\documentclass[10pt,conference]{IEEEtran}
\IEEEoverridecommandlockouts

\newif\ificseanonymous
\icseanonymousfalse

\usepackage[T1]{fontenc}
\usepackage{cite}
\usepackage{amsmath}
\usepackage{amssymb}
\usepackage{bbm}
\usepackage{graphicx}
\usepackage{xcolor}
\usepackage{booktabs}
\usepackage{tabularx}
\usepackage{multirow}
\usepackage{tablefootnote}
\usepackage{threeparttable}
\usepackage{pifont}
\usepackage{pbsi}
\usepackage{enumitem}
\usepackage[skins]{tcolorbox}
\usepackage{siunitx}
\usepackage[linesnumbered,ruled,vlined,noend]{algorithm2e}
\usepackage[caption=false,font=footnotesize]{subfig}
\usepackage{listings}
\usepackage{balance}
\usepackage{xurl}
\usepackage{comment}
\usepackage{etoolbox}
\usepackage[hidelinks]{hyperref}
\usepackage{tikz}
\usetikzlibrary{positioning, arrows.meta}

\setlist[itemize]{topsep=2pt, partopsep=0pt, itemsep=1pt, parsep=0pt}
\setlist[enumerate]{topsep=2pt, partopsep=0pt, itemsep=1pt, parsep=0pt}
\makeatletter
\patchcmd\algocf@Vline{\vrule}{\vrule \kern-0.4pt}{}{}
\patchcmd\algocf@Vsline{\vrule}{\vrule \kern-0.4pt}{}{}
\makeatother

\SetCommentSty{mycommfont}
\SetKwComment{Comment}{\ \#\ }{}
\SetKwProg{Func}{function}{}{end}

\tcbset{
  my box2/.style={
    enhanced,
    colframe=#1!80,
    colback=#1!5,
  },
}
\newtcolorbox{summary-rq}{
  my box2=black,
  boxrule=1pt,top=3pt,bottom=3pt,left=4pt,right=4pt
}

\AtBeginDocument{
  }
\providecommand{\Description}[1]{}

\title{File-Level Copying Is an Implicit Dependency in Open Source}

\ificseanonymous
  \author{\IEEEauthorblockN{Anonymous Author(s)}
  \IEEEauthorblockA{Anonymous Institution(s)}}
\else
  \author{
    \IEEEauthorblockN{Runzhi He}
    \IEEEauthorblockA{
      Peking University\\
      Beijing, China\\
      rzhe@pku.edu.cn}
    \and
    \IEEEauthorblockN{Audris Mockus}
    \IEEEauthorblockA{
      University of Tennessee, Knoxville\\
      Knoxville, TN, USA\\
      audris@utk.edu}
    \and
    \IEEEauthorblockN{Wenhao Yang}
    \IEEEauthorblockA{
      Peking University\\
      Beijing, China\\
      yangwh@stu.pku.edu.cn}
    \and
    \IEEEauthorblockN{Minghui Zhou\textsuperscript{*}}
    \IEEEauthorblockA{
      Peking University\\
      Beijing, China\\
      zhmh@pku.edu.cn}
  }
\fi

\begin{document}

\IEEEaftertitletext{\vspace{-24pt}}
\maketitle
\begingroup
\renewcommand{\thefootnote}{*}
\footnotetext{Corresponding author.}
\endgroup
\begin{abstract}
File-level copying is a widespread but ungoverned form of software reuse.
Copying files across repositories reduces \emph{supply-chain visibility}: it removes the four observable signals a package manager provides for a declared dependency (provenance, maintenance, security, and compliance) with no mechanism to restore them.
To characterize the scale and consequences of this unmanaged reuse, we present a mixed-method study of copying across the entire open-source ecosystem using World of Code (WoC). From a 0.1\% commit sample, we extract 690,500 copy events and retain 3,912 rationale-bearing copy commits for intent labeling. We show that the 13 axial copy forms, spanning vendored dependencies, hardware/driver synchronization, scaffolding, UI assets, and direct source-code reuse, are unreliable proxies for developer intent: among rationale-bearing commits, hardware/driver copies are predominantly fork-maintenance work (78\%), while dependency-vendoring copies more often signal upstream bypass (70\%) than offline availability.
These visibility gaps are form-specific: security and license risk concentrate in complementary copy forms.
Copied sources are frequently stale (median 155 days; 38.5\% over one year old) and seldom record a recoverable origin (4.3\% documented), let alone a checkable version (2.0\% versioned); even vendored copies record where they came from only 10\% of the time. Security risk concentrates in vendored dependencies: 17,314 CVE-risk copy commits in the full-WoC graph, 88\% in the dependency-vendoring form; 80\% score CVSS $\geq$7.0 and upstream-fix adoption is only 47\%--84\%. License risk concentrates in direct source-code reuse: 41,777 pre-validation candidates, 66\% in the source-code form, with 39 verified high-star violations ($\kappa=0.752$). Both risks reach packaged software and are invisible to dependency scanners operating on declared metadata alone. 

\end{abstract}


\section{Introduction}\label{chap:intro}

Software reuse frees developers from reinventing the wheel, accelerates release cycles, and forms an open source software supply chain of projects built on the work of others~\cite{Haefliger2008CodeRI}. Most supply-chain tooling focuses on explicit package dependencies: managers expose manifests, versions, lockfiles, and update channels that make it possible to reason about who depends on whom and where security or license obligations propagate~\cite{zimmermann2019small,pinckney2023flexible,DBLP:conf/icsr/ZeroualiCMRG18}.

File-level copying is a quieter form of reuse (what Mockus~\cite{mockus2019insights} termed a Type-II software supply chain), where the dependency is created by copying content rather than declaring a package reference. Developers copy source files, tests, workflows, generated assets, vendored libraries, and platform shims directly into repositories. Prior studies show this is common: Gharehyazie et al.~\cite{gharehyazie2017some} found that cross-project copies account for 10\%--30\% of GitHub clones; Hata et al.~\cite{Hata2021SameFD} found that more than 70\% of active non-fork repositories share files; and Jahanshahi et al.~\cite{jahanshahi2024beyond} showed, across billions of blobs in World of Code (WoC)~\cite{ma2021world}, that copy-based reuse is pervasive, that at least half of reused resources originate from small or medium projects, and that the infrastructure to govern such reuse is entirely absent. 

\textbf{File-level copying opens visibility gaps that open-source projects cannot ignore.}
For example, CVE-2016-9840 remained unpatched in PointCloudLibrary for nearly nine years. The vulnerable code had been ``cloned from zlib but did not receive the security patch''~\cite{pcl2025cve}, and because the copy carried no manifest entry or version field, no dependency scanner could link it to the advisory. Such orphan vulnerabilities recur at scale~\cite{DBLP:conf/icse/ReidJM22}, proving that
the invisibility is structural, with three compounding consequences.

\emph{First, copying removes supply-chain visibility without recording the loss.} Mockus~\cite{mockus2019insights} identifies \emph{visibility} (knowing who depends on what, at which version) and \emph{transparency} (making that dependency auditable) as core health properties of an OSS supply chain. A package manager provides both. A file copy transfers the same reuse but records none of it: no manifest entry, and no provenance, maintenance, security, or compliance metadata for any tool to act on. The visibility gap is universal across copies, while \emph{which dimension} dominates is form- and rationale-specific.

\emph{Second, the gap compounds silently over time.} With no automated drift signal, maintainers discover staleness by hand. \path{gdal}~\cite{repo_gdal}, packaged across 43 ecosystems, applied an upstream LibTIFF CVE fix in 3~days but still needed at least 61~days to reach its 20,768 dependent repositories. The delay exists only because the dependency is implicit; a pinned package would have proposed a bump on day one.

\emph{Third, the gap propagates downstream.} A project that copies a file and is itself packaged passes the same reduced visibility to every consumer, unnoticed. \path{discourse}~\cite{repo_discourse}, for example, ships an Apache-2.0 file inside a GPLv2 repository with no record of its origin, so downstream packagers inherit the license conflict without ever seeing it.

The central gap in existing knowledge is that copy studies detect shared artifacts but rarely recover the commit-level context in which files were copied, and without such contextual rationale, the maintenance implication of any given copy remains undetermined. A shared blob can indicate vendoring, fork synchronization, scaffolding, or generated output; each carries a different \textit{visibility-gap profile}, which determines what governance response, if any, is warranted. 
We target this gap with a mixed-method study of file-level copying in open source, mining 690,500 copy events from WoC~\cite{ma2021world} and combining multi-perspective signals with qualitative labeling of 3,912 copy commits.

Determining the appropriate governance response for a copy requires answers to three research questions in sequence.

\textbf{RQ1: What observable forms does file-level copying take?} A reviewer or tool first sees only the \textit{copy form}: the observable shape of a copy event, derivable from copied file types, destination paths, and repository context at scale.
\textit{Finding:} Copying spreads across 13 axial forms (11 substantive plus two disposition labels), from UI assets and scaffolding to vendored dependencies and device drivers, each with a distinct visibility-gap profile.

\textbf{RQ2: Why do developers copy files instead of depending on or abstracting them?}
Neither the literature nor existing tools explain why developers copy rather than declare a package dependency, 
so we label developer rationale from commit messages, copied paths, and linked pull-request or issue text.
\textit{Finding:} Among rationale-bearing commits, hardware/driver copies are predominantly fork-maintenance (78\%), while dependency-vendoring copies more often signal upstream bypass (70\%) than offline availability, showing that form and rationale diverge precisely where the governance stakes are highest.
With form capturing what a copy looks like and rationale explaining why it was made, we now measure how severely copying degrades supply-chain visibility, and where security and license risk concentrate.

\textbf{RQ3: How does file-level copying reduce supply-chain visibility, and where does that reduction concentrate risk?} Visibility (knowing who depends on what, at which version, through which update channel, with what maintenance activity and risk exposure) is what package managers provide and copying removes. We decompose RQ3 into four observable dimensions of this visibility gap: \textbf{RQ3a} Provenance (is the origin recorded and traceable?); \textbf{RQ3b} Maintenance (is the snapshot fresh, or has it silently drifted from upstream?); \textbf{RQ3c} Security (do copies carry known-vulnerable code, and do upstream fixes reach them?); and \textbf{RQ3d} Compliance (is license context preserved?), stratified by copy form across 690,500 events and interpreted through the rationale associations from RQ2.
\textit{Finding:} Copied sources are often undocumented and stale; CVE risk concentrates in vendored dependencies (88\%, 80\% high-severity); license risk concentrates in source-code reuse (66\%); both reach packaged software invisible to manifest-based tooling. The result is a form-aware, rationale-informed \emph{visibility-gap profile} indicating \textit{which dimension} is most affected and \textit{where risk concentrates}, 
suggesting that a single clone-detection rule is poorly matched to this variation. 

This paper makes five contributions:
\vspace{-.05in}
\begin{itemize}
    \item A reusable copy-commit tracing and source-attribution pipeline. 
      As the technical prerequisite for this work, it engineers copy-seed detection over billions of blobs, 
      event reconstruction, and source attribution (89.5\% accuracy), producing a 690,500-event enriched dataset 
      as a reusable artifact for future supply-chain studies.
    \item A large-scale, commit-level taxonomy of copy forms across 690,500 copy events, built on a dual-annotated 13-category axial codebook
    and applied at scale by an LLM labeler, spanning uses from vendored dependencies to direct source-code reuse.
    \item A rationale codebook for why developers copy files, grounded in 3,912 labeled commits
    , showing that observable copy forms only partially explain developer intent, with sharp form--rationale associations diagnostic for tool design.
    \item Quantification of the supply-chain visibility loss copying creates across four observable dimensions (provenance, maintenance, security, compliance), stratified by copy form and interpreted through rationale.
    \item Evidence that these visibility gaps persist even in package-distributed software, where manifest-based tooling cannot see them, and the resulting requirements for an implicit dependency manager to restore visibility for copied files: blob-level CVE scanning for vendored copies, license-pair checks for source-code reuse, and upstream-trailer linting for fork maintenance.
\end{itemize}

\section{Background}
\subsection{Software Supply Chains}
Software supply-chain (SSC) research models software as a network of upstream components, maintainers, packages, build systems, and downstream consumers. Prior work~\cite{mockus2019insights} distinguishes two supply-chain types in OSS: Type-I (explicit package dependencies managed through registries and lockfiles) and Type-II (file-level copying, where content is transferred without a manifest entry). 
Most SSC research (transitive dependency networks, registry attack surfaces, version-resolution studies) addresses Type-I.
Package managers make Type-I dependencies observable: manifests and lockfiles record names and versions, registries expose releases, and solvers reason about updates, compatibility, and risk. 
This observability has enabled empirical work on
 package ecosystem structure~\cite{zimmermann2019small}, dependency resolution~\cite{pinckney2023flexible}, automated dependency updates~\cite{DBLP:journals/tse/HeHZZ23}, vulnerabilities and malicious packages~\cite{ohm2020backstabber}, license compliance~\cite{cui2023empirical,xu2023understanding}, and maintenance pressure from dependency change~\cite{DBLP:conf/icsr/ZeroualiCMRG18,wu2016exploratory}. Kula et al.~\cite{DBLP:journals/ese/KulaGOII18} found that 81.5\% of systems leave vulnerable dependencies un-updated; studies confirm weeks-to-months lags in vulnerability-fix adoption and widespread downstream propagation~\cite{DBLP:journals/smr/ZeroualiMGDCR19,chinthanet2021lags,DBLP:conf/icse/LiuCF00022,DBLP:journals/cacm/Cox19}. File-level copies carry the same obligations but generate no manifest entry and no registry record, so none of these tools and studies can detect or track them.

This paper studies Type-II, the implicit dependency created when a file is copied without a package reference. 
We use \emph{visibility} broadly~\cite{mockus2019insights}: knowing who depends on what, at which version, through which update channel, with what maintenance activity and risk exposure. Provenance (where an artifact originated) is one narrow facet; a copied file can carry a perfect source URL yet still be a stale, CVE-laden snapshot with no update mechanism.

\subsection{Copying Practices and Risks}

Prior work establishes that copy-based reuse is widespread, spanning file types well beyond source code~\cite{gharehyazie2017some,Hata2021SameFD,jahanshahi2024beyond}. Focused studies explain why copies appear: vendored libraries remain embedded for years~\cite{sync2019javascript,wang2021hero}; fork descendants must manually propagate fixes~\cite{jones2020deploying,businge2022reuse}; and configuration files are copied as starting points~\cite{cardoen2024empirical,saroar2023developers,oumaziz2019handling,mcintosh2014collecting,zhu2025empirical}.

Copying separates reuse from the metadata that normally helps maintain it, creating security, compliance, and maintenance risks. 
For \emph{security}, copied code may stop receiving upstream patches: Wyss et al.~\cite{wyss2022what} found 6,292 npm ``shrinkwrapped clones'' that embed another package without a dependency declaration, 207 of them carrying vulnerabilities invisible to \texttt{npm audit}; Reid et al.~\cite{DBLP:conf/icse/ReidJM22} 
~\cite{reid2023uvhistory} 
detected such orphan vulnerabilities (CVEs fixed upstream but still present in copies) at WoC scale. Ramkisoen et al.~\cite{ramkisoen2022pareco} measured a median patch lag of 27 weeks across copied-file variants; Jahanshahi et al.~\cite{jahanshahi2025cracks} traced the same risks into LLM pre-training corpora. 
For \emph{compliance}, SCA tools reason only over declared dependencies, so copying can strip license context entirely: Golubev et al.~\cite{golubev2020study} estimated 9.4\% of copied Java blocks potentially violate their source licenses, and Jahanshahi et al.~\cite{jahanshahi2025integrity} showed that 39.4\% of project pairs in the reuse network are at noncompliance risk while only 2.43\% of the reuse is discoverable through dependency analysis. Copying also accrues \emph{maintenance} and bloat overhead, carrying unused dependencies and generated artifacts into the destination~\cite{liu2025detecting}.

\subsection{Gaps}
\vspace{-.04in}
Existing work leaves two gaps this paper addresses. First, while Jahanshahi et al.~\cite{jahanshahi2024beyond} asked developers \emph{why} they copy (eight motivational themes from 374 surveyed), those survey categories are too coarse to determine the visibility dimension a copy most affects: knowing a copy was made for ``functionality'' does not reveal whether a vendored file was an intentional upstream bypass (a security-visibility gap) or an offline-availability pin (a freshness gap). The prior work reviewed above operates on blob identity, project-pair relationships, or self-reported motivation~\cite{jahanshahi2024beyond,jahanshahi2024dataset,wyss2022what,DBLP:conf/icse/ReidJM22,reid2023uvhistory,jahanshahi2025cracks,jahanshahi2025integrity,jahanshahi2024license}; none recovers the \emph{commit-level} context (copy form, rationale, provenance, freshness) needed to determine \emph{which} visibility dimension a \emph{specific} copy affects. Second, existing risk analyses treat copied code as one undifferentiated class, but our results show CVE risk concentrates in dependency-vendoring (88\%) while license risk concentrates in source-code reuse (66\%), so form-blind rules simultaneously over-alert on fork synchronization and under-specify the gaps that matter. We address both gaps by studying copying through three linked views: observable forms (RQ1), gap-discriminating commit-level rationales (RQ2), and downstream visibility gaps stratified by form (RQ3).

\section{Methodology}
\label{sec:analysis_methods}

Two commitments shape our methodology. \textbf{Scale and provenance:} studying file-level copying across the open-source ecosystem requires a corpus that links file content to \emph{every} commit and project containing it, so we build on WoC~\cite{ma2021world}, whose blob-level maps provide cross-project provenance beyond the coverage of any single forge. \textbf{Mixed methods:} observable copy forms and downstream visibility gaps 
are measurable quantitatively at scale, whereas developer rationale must be interpreted from commit text, so we pair large-scale mining with qualitative coding. 
\vspace{-.1in}
\subsection{Data Construction}
\label{sec:data_construction}
\vspace{-.05in}

\textbf{Copy-seed identification.}
Unlike token-, AST-, or PDG-based clone detection~\cite{roy2009comparison,svajlenko2017fast,semura2017ccfindersw}, which finds similar fragments but does not scale to billions of blobs, we detect copies by exact SHA-1 blob identity.
 A file copy is therefore detectable only when the same blob content recurs across projects. From WoC's \texttt{bb2cf} map (blob $\rightarrow$ the commits, paths, and parent blobs that introduce it) we mark a blob as a \emph{copy seed} when it appears in more than one tree \emph{and} more than one project, after discarding merge commits and rebased or amended duplicates that would otherwise inflate the count with artifacts rather than genuine copies. This yields roughly 2.2 billion copy-seed blobs (7.4\% of all blobs in WoC), which we pack into a Bloom filter (0.1\% false-positive rate) so that any commit can be tested for copy membership in $O(1)$ without re-querying WoC.

\textbf{Copy-commit detection.} Enumerating every commit that touches a seed would bias the corpus toward already-copy-heavy blobs, so we instead draw an approximately uniform 0.1\% Bernoulli sample over commits (\texttt{c2fbb}) and test each one. A sampled commit is a \emph{copy commit} when a changed file's new blob is in the Bloom filter \emph{and} that blob's earliest author time (\texttt{b2fa}) precedes the commit, so the content demonstrably existed earlier elsewhere. We skip placeholder and metadata filenames and IDE/OS/SCM paths, but deliberately \emph{keep} vendored directories such as \path{node_modules/} and \path{third_party/}: vendoring is a valid copy form, and filtering it would bias the form taxonomy.

\textbf{From commits to events.} A single logical copy often surfaces as several near-duplicate commits: rebases rewrite SHAs and timestamps, and when an upstream branch is later deleted WoC cannot link the true-earliest commit, so a downstream repository that preserved the change can \emph{appear} earliest. We therefore collapse near-duplicate commits that represent the same logical change into \textbf{events}, taking the earliest committer time and unioning the projects. One event is one logical change; its earliest member is the creation/source and the rest are copies. Defining events rather than counting commits is what makes ``earliest~$=$~source'' well-posed.

\textbf{Noise control.} Ecosystem-scale WoC 
carries systematic artifacts, removed by a deliberate filter chain: WoC bad-object lists of corrupt records (authors, commits, blobs, trees, and projects); residual paths and pre-Git-era commits left by SVN-to-Git imports; and grafted or rewritten history, detected when a candidate's parent carries a \emph{later} timestamp than the candidate itself~\cite{flint2022pitfalls}.
After these filters the pipeline retains 690,500 copy events. We enrich each with WoC and GHArchive~\cite{gharchive}, a public archive of GitHub's public event timeline, adding commit metadata, repository statistics, pull-request and issue links, and file diffs; each RQ then draws its population from this corpus, as detailed in the per-RQ methods below.

\textbf{Source-origin attribution.} Attribution has two parts. The source \emph{commit} is the earliest event that survives the filters above; its later events are copies. The source \emph{project} is harder, because WoC's raw source pointer is the \emph{nearest} copy (typically a mirror or fork), not the true origin. We therefore score every candidate project $p$ for a copied blob and take the argmax:
\begin{equation*}
\label{eq:attribution}
\begin{aligned}
\mathrm{score}(p) ={}& 0.35\,\mathrm{sim}(p,a) + 0.25\,\mathrm{pop}(p) + 0.20\,\mathrm{com}(p) \\
 &{} + 0.20\,\mathbbm{1}[\,p \in G\,] + 0.50\,\mathrm{temp}(p)
\end{aligned}
\end{equation*}
where $\mathrm{sim}(p,a)$ is the fuzzy similarity between the project owner and the commit author $a$; $\mathrm{pop}(p)=\tfrac{1}{2}(\log\text{forks}+\log\text{stars})$ and $\mathrm{com}(p)=\tfrac{1}{2}(\log\text{issues}+\log\text{PRs})$ are $\log$-scaled GHArchive popularity and community-activity counts; $\mathbbm{1}[\,p \in G\,]$ marks membership in a curated set $G$ of ground-truth upstream repositories (Repology $\cup$ CVE/VFC records); and $\mathrm{temp}(p)=\log(\text{age of }p\text{ at the commit})$, or $-1$ when $p$ was created after the commit and so cannot be the source. We set these weights by hand on small labeled samples rather than fitting them, keeping the score interpretable; the rule-based design deliberately trades marginal ranking accuracy for an attribution that scales to billions of blobs while staying fully reproducible and auditable from WoC and GHArchive statistics alone.

\textbf{Evaluation.} One author manually validated 502 attributions stratified by programming language: accuracy was \textbf{89.5\%}, with errors dominated by upstreams not indexed by WoC (25), unreliable popularity signals for low-activity projects (12), anomalous commit timestamps that make a downstream repository appear earliest (11), and ambiguous cherry-pick or minified histories (5).
\vspace{-.1in}
\subsection{RQ1: Copy-Form Taxonomy}
\label{sec:rq1_method}
\vspace{-.05in}
We developed the RQ1 copy-form taxonomy through inductive thematic analysis. Two coders open-coded a stratified sample of copy commits by file type and repository context, then axially grouped the open codes into 13 axial categories. 

Several axial categories are relational or intent-bearing (e.g., \textit{upstream sync and forking}, \textit{project migration and refactoring}) and cannot be detected from path and file-type signals alone. We apply this 13-category axial codebook at scale using a deterministic LLM labeler (\texttt{gemini-3-flash}, temperature 0.0) to all 690,500 copy events, producing the copy-form distribution in Table~\ref{tab:taxonomy} and the RQ1 axis of the RQ1--RQ2 association (Figure~\ref{fig:rq1_rq2_heatmap}).
All form-stratified RQ3 analyses use the validated 13-axial LLM labels, not a separate path/extension classifier. For the 0.1\% commit sample we stratify each RQ3a (provenance) and RQ3b (maintenance) measure by the full-scale axial label of its event (Section~\ref{sec:rq1_method}). For the full-WoC RQ3c/d risk commits, which lie outside the sample, we apply the same deterministic LLM labeler 
to the same inputs used for RQ1--RQ2 (the flagged files, their content, and the commit message) for each CVE- and license-risk commit, so the codebook's validation $\kappa$ (Section~\ref{sec:rq1_method}) carries over to these labels; we group the 13 codes into broader source-code and vendoring categories for the compact risk tables. 

\textbf{Labeler agreement.} We assessed the reliability of the \textit{axial} coding on a stratified sample of 257 copy commits annotated independently by two coders using the 13-category codebook. Inter-annotator agreement was strong~\cite{landis1977measurement}: raw agreement 86.4\%, Cohen's~\cite{cohen1960coefficient} $\kappa=0.8452$, macro $\kappa=0.7989$, with per-category $\kappa$ ranging from 0.497 (ambiguous / unknown) to 0.932 (dependency vendoring). Against the human consensus, an LLM labeler reproduced the axial labels at 82.4\% agreement ($\kappa=0.7992$, per-category up to 0.876 for coincidental and generated). 
\vspace{-.1in}
\subsection{RQ2: Rationale Labeling}
\label{sec:rq2_method}
\vspace{-.05in}
Because most copy commits do not explain the rationale, RQ2 focuses on an intent-rich subset rather than estimating prevalence. \textbf{Candidate selection.} Heuristic signals (linked PR, non-trivial issue body, rationale-bearing message keywords) followed by a rule-based boilerplate filter retain 66,520 candidates from the 690,500 events. An LLM relevance pass keeps the 3,912 commits on which repeated judgments agree ($\geq$0.9) as high-confidence rationale-bearing commits.

\textbf{Codebook development.} The codebook was built manually using qualitative content analysis with \textit{provisional coding}~\cite{saldana2013coding}.
We seeded a provisional ``start list'' of codes synthesized from prior studies of copy-based reuse~\cite{sync2019javascript,businge2022reuse,jones2020deploying}, which two annotators then refined and extended through constant comparison while labeling commits, producing the nine production codes in Table~\ref{tab:rq2_codebook}. 
A 300-commit check disjoint from the codebook-building sample found no new recurring rationale among the unmapped commits, consistent with codebook saturation for recurring copy rationales. 
For interpretation, the production codes group into five structural \textit{boundaries}: Lineage (CF), Dependency (UB, OV), Execution (HO, PF), Architecture (MD), and Institutional/platform (PS, TS).

\textbf{Scaling and reliability.} With the codebook fixed, we use a deterministic LLM labeler (\texttt{gemini-3-flash}, temperature 0.0) to apply the final prompt and codebook to all 3,912 commits. The same LLM labeler produces the RQ1 axis of the RQ1--RQ2 association. We compute the cross-table (Figure~\ref{fig:rq1_rq2_heatmap}) over the 3,494 commits that remain after excluding the 252 \textsc{EXCL} cases and the 166 \textit{coincidental\_and\_generated} cases that are not \textsc{EXCL} (an additional 35 commits carry both labels and are already counted in the 252).
To assess rationale-code reliability, two coders independently labeled a stratified 300-commit sample, followed by adjudication. On in-taxonomy items (excluding \textsc{EXCL}), raw agreement was 87.9\%, Cohen's $\kappa=0.765$ ($n=231$); including the \textsc{EXCL} gate, raw 76.3\%, $\kappa=0.650$ ($n=300$). Residual divergence concentrates at the in-scope/out-of-scope gate ($\kappa=0.477$); full inter-rater detail appears in 
Section~\ref{sec:threats}.

\subsection{RQ3: Visibility-Gap Measures}
\label{sec:rq3_method}
The four RQ3 dimensions draw on overlapping but distinct populations. RQ3a, RQ3b, and RQ3d all derive from the copy-event corpus (RQ3a and RQ3b on the 690,500-event commit sample, RQ3d on project pairs from the same corpus), while RQ3c is seeded separately, from CVE/VFC vulnerable blobs (below), because the rare security events are too sparse to catch at a 0.1\% sampling rate. Denominators are not comparable across dimensions.

\textbf{Package-repository reach.} To test whether copy-based visibility gaps reach packaged software, we map destination projects to Repology~\cite{repology}, a cross-ecosystem package aggregator (221,489 records over 174,290 WoC projects; median 1 mapped ecosystem per project, p99=28); a repository mapped to many ecosystems participates in package distribution at scale, though Repology does not confirm that a specific copied file ships in a released artifact. Where we report package reach (RQ3b, RQ3d) we apply de-noise filters that exclude backport/patch-porting and mirror/import context, same-owner or near-identical source--destination pairs, and non-production files (test fixtures, generated files).

\textbf{RQ3a (Provenance).}
We measure provenance by what a future maintainer could \emph{act on}, using a deterministic detector over each event's recorded text (commit message, pull-request and issue text) and the copied-file content in the enrichment diff. 
We count only a \emph{pointer}: a recorded identifier that resolves to a specific upstream (a source URL, owner/repo slug, commit hash, submodule, manifest source field, or explicit ``derived from $\langle$named project$\rangle$'').
We report two \emph{nested} levels: \textit{documented} provenance means a pointer is recoverable, so a maintainer can learn where the file came from, and \textit{versioned} provenance (a subset) additionally pins an upstream version (commit hash, submodule SHA, version-qualified URL, or manifest version), so a maintainer can tell whether the copy is stale; the \textit{visibility gap} is the complement. Because the detector sees only process text and a truncated content sample, the rates are a conservative lower bound; an author audit of 100 stratified events (50 documented, 50 gap) found 89\% construct alignment (documented precision 90\%; 88\% of gap cases carried no recoverable pointer). A \textit{pull-request link} marks events that passed through a reviewable channel, reported separately as process visibility, and repository stars serve as a review-capacity proxy. We define \textit{no recorded maintenance memory} as cross-project events with at least three co-occurring gap signals (snapshot lag $>$1y, age span $>$3y, copy ratio $>$0.9, batch size $\geq$10, attention gap, package-mapped destination, or high-fanout source) and no PR link, documented provenance, or lifecycle text. We test visibility--gap associations with Mann--Whitney $U$ (Cliff's $\delta$) and Fisher's exact tests (odds ratios); full-WoC provenance rates (RQ3c/d) use commit messages only, and all associations are observational.

\textbf{RQ3b (Maintenance).} For each copy event, \textit{snapshot lag} is the destination commit author time minus the latest copied-blob creation time. In principle this uses all copied blobs, but large commits can carry thousands of blobs, so for efficiency we sample up to ten when an event copies ten or more files and take the newest sampled blob. Because we cannot establish with certainty where each file was copied from, snapshot lag is an \emph{estimate} of the technical lag to the immediate copy source rather than the true origin. \textit{Age span} uses the earliest sampled blob-created time, measuring historical audit depth. We report both but use snapshot lag as the primary technical-lag measure.

\textbf{RQ3c (Security).} RQ3c draws on a separate, seeded dataset. From the vulnerability fixing commits (VFC) dataset of Nguyen et al.~\cite{nguyen2025mappingvfc} we take a CVE vulnerable/fixed blob table (108,135 changed-file rows over 15,732 CVEs, with pre-fix (\path{old_blob}) and post-fix (\path{new_blob}) blobs and CVSS~\cite{cvss31} metadata) and seed from the vulnerable \path{old_blob}s, applying the copy-detection method of Section~\ref{sec:data_construction} to locate every commit that copied one. This yields 17,314 CVE-risk copy commits. Repair observability is bracketed by a strict lower bound (the exact fixed blob later appears in the destination) and a permissive upper bound (the vulnerable blob is later edited away, traced through \texttt{obb2cf}); we read the pair as an interval, not a patch rate. \textit{Repair delay} measures the time from upstream fix to downstream fix where both are observable.

\textbf{RQ3d (Compliance).} RQ3d uses a full-scale but heavily pruned dataset: a project-to-project copy mapping over the corpus, restricted to pairs where \emph{both} repositories carry at least 10 stars (the coverage of SEART GitHub Search~\cite{dabic2021sampling}, whose license metadata we rely on). Joining these pairs to repository licenses and a directional SPDX-style compatibility matrix following Wu et al.~\cite{wu2021licenseSelection} gives 41,777 pre-validation incompatible candidates. We manually review the high-star subset (destination $\geq$1,000 stars) to separate plausible violations from false positives (same third-party dependency in both projects, reversed dependency direction, dual licensing, generated files, etc.). A second independent audit re-judges all 326 high-star rows at the \emph{file level}: comparing the historically copied blob's own license (read from WoC) against the destination license, treating documentation, manifests, and configuration as mere aggregation, and producing a binary confirmed/false-positive label that corrects the repository-level over-approximation.

\section{Results}
\label{sec:results}

\subsection{RQ1: Copy Forms}
\label{sec:rq1_results}
\vspace{-.02in}
Table~\ref{tab:taxonomy} defines the 13 axial copy-form categories (Section~\ref{sec:rq1_method}). The defining feature of the distribution is its shape: copying extends well beyond direct application-source reuse. No single form dominates the substantive distribution: UI assets and media (14.6\%), configuration and infrastructure (11.4\%), and scaffolding or templates (8.4\%) lead, while the forms most associated with reuse obligations remain modest: dependency vendoring (7.0\%) and hardware/drivers (2.2\%). The single largest bucket overall is \textit{coincidental\_and\_generated} (28.9\%), capturing detection false positives (generator output, common defaults, coincidental convergence) and excluded from substantive form analysis.
A reuse-governance model scoped to source clones alone would therefore miss most of the copying that actually happens. We compute the RQ1--RQ2 association below on the 3,912 rationale-bearing commits.

\begin{table*}[t]
\centering
\caption{RQ1 copy-form taxonomy: the 13 axial codebook categories (Section~\ref{sec:rq1_method}), with counts and shares over all 690,500 LLM-labeled copy events.
}
\label{tab:taxonomy}
\vspace{-.1in}
\footnotesize
\begin{tabularx}{\textwidth}{@{}l>{\raggedright\arraybackslash}X rr@{}}
\toprule
\textbf{Copy form} & \textbf{Description} & \textbf{Count} & \textbf{Share} \\
\midrule
UI Assets \& Media & Visual or media resources shown to end users, plus presentational code (CSS, HTML templates, UI markup) that defines look-and-feel; files are rendered rather than executed. & 100,610 & 14.6\% \\
Configuration \& Infrastructure & Build files, CI/CD workflows, container and deployment specs, and tooling conventions that govern how software is built, tested, and deployed rather than runtime logic. & 78,739 & 11.4\% \\
Scaffolding \& Templates & Initializing a new project or module by copying a framework skeleton, generator output, or starter template; the copied files form the repository's initial structure. & 58,198 & 8.4\% \\
Upstream Sync \& Forking & Copying while maintaining a fork, synchronizing or backporting from an upstream project, cherry-picking, or mirroring a repository under a fork-parent relationship. & 55,854 & 8.1\% \\
Course Materials & Copied content that itself functions as instructional material: course exercises, homework starter code, tutorial files, or assignment solutions. & 51,587 & 7.5\% \\
Dependency Vendoring & In-tree inclusion of a third-party library, SDK, or plugin directly in the source tree, bypassing the package manager; the artifact is externally authored and maintained separately. & 48,283 & 7.0\% \\
Logic \& Algorithms & Ad-hoc copying of a few files with self-contained algorithms, utilities, or business logic from an external source (an isolated snippet, not a whole library). & 27,685 & 4.0\% \\
Data \& Datasets & Structured data, databases, corpora, or reference collections copied as input to be processed rather than code executed or assets displayed. & 23,487 & 3.4\% \\
Hardware \& Drivers & Firmware, hardware drivers, kernel device-tree files, or hardware-specific scripts that enable specific physical hardware; typical of embedded, OEM-kernel, and IoT projects. & 14,937 & 2.2\% \\
Project Migration \& Refactoring & Deliberate reorganization across an author's or organization's own repositories: extracting a module, splitting a monorepo, or reviving an abandoned project. & 14,409 & 2.1\% \\
Test Artifacts & Test cases, suites, fixtures, and test utilities executed by a test runner to verify behavior rather than provide application functionality. & 13,984 & 2.0\% \\
\midrule
Coincidental \& Generated & Blob-similarity flagged without intentional copying: generator output, common defaults, or coincidental convergence (noise / false positive). & 199,212 & 28.9\% \\
Ambiguous / Unknown & Evidence insufficient to assign any other category with reasonable confidence; a residual category of last resort. & 3,515 & 0.5\% \\
\bottomrule
\end{tabularx}
\vspace{-.1in}
\end{table*}
\vspace{-.05in}
\subsection{RQ2: Developer Rationales}
\label{sec:rq2_results}
\vspace{-.02in}
Of 3,912 labeled copy commits, 3,660 enter the main RQ2 analysis; the remaining 252 are labeled \textsc{EXCL} (commits whose evidence describes an ordinary project change, such as a routine edit, merge, or refactor, rather than a copy-specific rationale) and are retained for auditability but excluded from the distribution.
Table~\ref{tab:rq2_codebook} defines the nine codes and reports their distribution over this intent-rich subset (not a prevalence estimate over all 690,500 events; Section~\ref{sec:rq2_method}). Cross-project patch porting dominates (CF, 57.30\%): many rationale-bearing copies are synchronization work across forks or downstream distributions, e.g.\ a kernel fork porting an upstream ext4 fix (``correctly migrate a file\dots\ commit 8974fec7\dots\ upstream''). Upstream bypass is the second rationale (UB, 23.17\%): developers embed a dependency because upstream does not satisfy immediate needs. Together these two account for 80\% of labeled commits; the remaining rationales mark structurally distinct boundaries, from offline vendoring (OV, ``so I don't have to publish to npm'') to transitional scaffolding (TS, ``temporary changes\dots\ to run against [a] devel COPR fork''). A vendored-looking file may patch an upstream defect, and a source copy may be disposable scaffolding. Unlike the survey-based motivational themes of Jahanshahi et al.~\cite{jahanshahi2024beyond}, our commit-level codes separate upstream bypass (UB) from offline vendoring (OV), the distinction that determines the visibility gap, and surface cross-project patch porting (CF, 57\%) as maintenance work, not a reason to copy.

\begin{table*}[t]
\centering
\caption{RQ2 rationale codebook over the 3,660 rationale-coded commits (3,912 minus 252 \textsc{EXCL}).
}
\label{tab:rq2_codebook}
\vspace{-.1in}
\footnotesize
\begin{tabularx}{\textwidth}{@{}ll l >{\raggedright\arraybackslash}X rr@{}}
\toprule
\textbf{Code} & \textbf{Rationale} & \textbf{Boundary} & \textbf{Description} & \textbf{Count} & \textbf{Share} \\
\midrule
CF & Cross-project patch porting & Lineage & Downstream fork or branch ports/backports upstream work due to structural divergence. & 2,097 & 57.30\% \\
UB & Upstream bypass & Dependency & Destination patches or embeds a dependency because upstream is broken, slow, or unresponsive. & 848 & 23.17\% \\
PF & Performance/footprint & Execution & Copying removes abstraction overhead or reduces binary size. & 191 & 5.22\% \\
MD & Module decoupling & Architecture & Copying repairs internal module boundaries or publishes a component independently. & 178 & 4.86\% \\
OV & Offline vendoring & Dependency & File kept in-tree for availability, reproducibility, or policy, without a local correctness patch. & 128 & 3.50\% \\
HO & Hardware/platform quirks & Execution & Copy handles hardware anomalies, compiler bugs, or ABI differences. & 102 & 2.79\% \\
PS & Permanent shim & Institutional & Durable in-tree shim imposed by platform, organization, or compatibility policy. & 61 & 1.67\% \\
TS & Temporary scaffolding & Institutional & Explicitly transitional copy during an active migration or staged replacement. & 48 & 1.31\% \\
\midrule
ED & Educational reuse & --- & Copy used for assignments, examples, or learning. & 7 & 0.19\% \\
EXCL & Excluded & --- & Evidence describes an ordinary project change (routine edit, merge, refactor), not a copy-specific rationale. & 252 & --- \\
\bottomrule
\end{tabularx}
\vspace{-.01in}
\end{table*}

The five structural boundaries carry distinct maintenance profiles, reflecting that developers are doing fundamentally different things: Lineage copies are deep-history synchronization work while Architecture copies are fresh restructuring, and Dependency copies are handled very differently from Execution or Institutional ones. We quantify these boundary-specific profiles with the RQ3b maintenance measures (Section~\ref{sec:rq3_implications}).

\begin{figure}[tbp]
\vspace{-.05in}
\centering
    \includegraphics[width=\columnwidth]{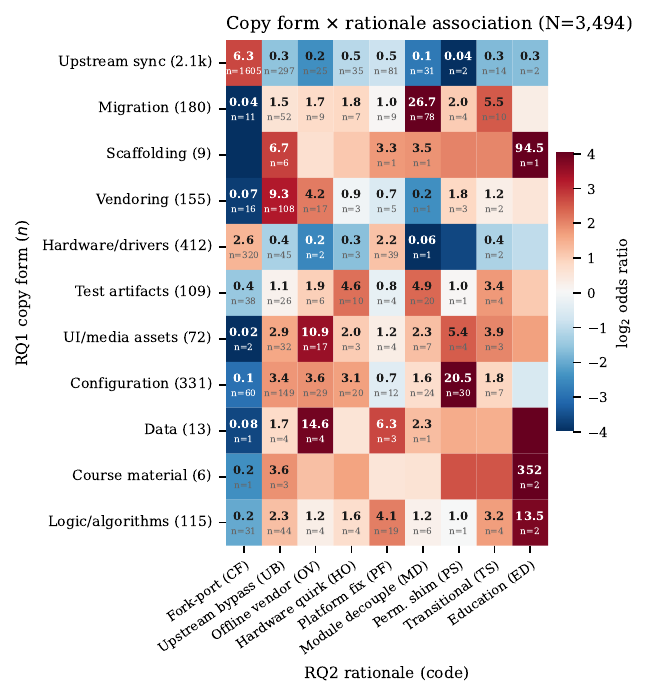}
    \vspace{-22pt}
    \caption{RQ1 copy form vs.\ RQ2 rationale (3,494 in-taxonomy commits)
}
    \label{fig:rq1_rq2_heatmap}
    \Description{Heatmap of RQ1 copy forms by RQ2 rationale categories: each cell shows the one-vs-rest odds ratio over the raw count, colour encodes the log2 odds ratio, and per-form and per-code totals appear in the axis labels.}
\vspace{-.08in}
\end{figure}

\paragraph{RQ1--RQ2 Association.}
Figure~\ref{fig:rq1_rq2_heatmap} cross-tabulates the LLM-annotated 13-axial copy form and the RQ2 rationale on the 3,494 in-taxonomy commits (each cell: one-vs-rest odds ratio over count; colour: log$_2$ odds ratio). Axial form and rationale are associated but not interchangeable. The strongest diagonal signal is \textit{upstream\_sync\_and\_forking} $\rightarrow$ CF: 76.7\% of upstream-sync copies are cross-project patch porting (OR $6.3\times$), indicating that this axial category is the observable surface of fork maintenance. \textit{Hardware \& drivers} copies also map predominantly to CF (77.7\%, OR $2.6\times$), not to the hardware-quirk rationale (HO, 0.7\%); the path signal ``kernel driver'' is therefore a fork-maintenance surface even when the axial label is content-correct. \textit{Dependency vendoring} is the cleanest intent mismatch: 69.7\% of vendored copies are upstream-bypass cases (OR $9.3\times$), and only 11.0\% are offline-vendoring (OV); the path ``vendor/'' is therefore a stronger upstream-bypass signal than an offline-pin signal. Two source-code-type axial categories carry their own rationale signatures: \textit{project\_migration\_and\_refactoring} is dominated by module decoupling (43.3\%, OR $26.7\times$), and \textit{logic\_and\_algorithms} concentrates in upstream bypass (38.3\%) and patch porting (27.0\%). Configuration/infrastructure copies associate most strongly with permanent organizational shims (PS, OR $20.5\times$) and upstream bypass (UB, 45.0\%). These form-rationale mismatches are consequential for governance and shape the form-specific responses in the Discussion (Section~\ref{sec:discussion}).

\vspace{-.05in}
\begin{summary-rq}
\textbf{RQ1\,\&\,RQ2 summary.} Copying spans eleven substantive forms, but form does not determine intent: among rationale-bearing commits, hardware/driver copies are 78\% fork-maintenance; vendoring copies are 70\% upstream bypass. The rationale, not the form, determines which visibility dimension dominates.
\end{summary-rq}

\subsection{RQ3: Visibility Gaps}
\label{sec:rq3_impact_overview}
\vspace{-.06in}
We examine supply-chain visibility loss across four observable dimensions: provenance (RQ3a) $\rightarrow$ maintenance (RQ3b) $\rightarrow$ security (RQ3c) $\rightarrow$ compliance (RQ3d). Provenance comes first because it conditions the rest: an unrecorded copy cannot be audited, version-tracked, or reliably attributed. Each dimension is sharply form-specific: security and compliance concentrate in complementary forms (dependency vendoring vs.\ source-code reuse), while provenance and maintenance gaps spread more diffusely across ad-hoc and peripheral copies. All reach package-distributed repositories outside any manifest. Different datasets underlie different dimensions (Section~\ref{sec:rq3_method}), so denominators are not comparable across them.

\vspace{-.09in}
\subsection{RQ3a: Provenance}
\label{sec:rq3_attention}
\vspace{-.05in}
Over the 690,500-event commit sample, only 4.25\% of events carry \textit{documented} provenance (a recorded pointer the maintainer could follow) and only 1.97\% carry \textit{versioned} provenance, a pointer that also pins an upstream version, leaving 95.75\% with no recorded provenance at all. Provenance is sharply form-specific (Table~\ref{tab:rq3a_rq3b}). Hardware/driver copies are the best documented (21.9\% documented, 20.5\% versioned), almost entirely through upstream-commit references in fork maintenance, followed by dependency vendoring (10.5\%) and upstream sync/forking (10.3\%). Dependency vendoring is the instructive case: only 10.5\% \emph{record} a source pointer and 2.3\% a version, so most in-tree third-party code carries no recoverable pointer and a maintainer can rarely learn \emph{which} upstream or version it tracks. The visibility gap is widest exactly where origin matters most: ad-hoc logic/algorithm copies (97.8\% undocumented), data (99.4\%), and course materials (99.3\%), where no vendoring or fork convention carries the origin.

\begin{table*}[t]
\centering
\caption{Provenance (RQ3a) and maintenance (RQ3b) by RQ1 copy form over the 690,500-event sample. 
}
\label{tab:rq3a_rq3b}
\vspace{-.1in}
\footnotesize
\begin{tabular}{lrrrrr}
\toprule
& \multicolumn{2}{c}{\textbf{Provenance (RQ3a)}} & \multicolumn{3}{c}{\textbf{Maintenance (RQ3b)}}  \\
\cmidrule(lr){2-3} \cmidrule(lr){4-6}
\textbf{Copy form} & \textbf{Documented} & \textbf{Versioned} & \textbf{Events} & \textbf{Median days} & \textbf{$>$1yr} \\
\midrule
UI Assets \& Media & 3.89\% & 0.52\% & 100,610 & 218.4 & 40.0\% \\
Configuration \& Infrastructure & 3.73\% & 1.26\% & 78,739 & 90.7 & 31.2\% \\
Scaffolding \& Templates & 6.15\% & 0.32\% & 58,198 & 229.2 & 40.1\% \\
Upstream Sync \& Forking & 10.30\% & 8.85\% & 55,854 & 6.8 & 23.4\% \\
Course Materials & 0.70\% & 0.05\% & 51,587 & 268.7 & 43.7\% \\
Dependency Vendoring & 10.46\% & 2.25\% & 48,283 & 107.1 & 31.4\% \\
Logic \& Algorithms & 2.16\% & 1.08\% & 27,685 & 141.0 & 34.1\% \\
Data \& Datasets & 0.64\% & 0.35\% & 23,487 & 79.2 & 33.9\% \\
Hardware \& Drivers & 21.88\% & 20.47\% & 14,937 & 155.2 & 37.0\% \\
Project Migration \& Refactoring & 3.72\% & 2.49\% & 14,409 & 3.0 & 8.3\% \\
Test Artifacts & 4.23\% & 3.38\% & 13,984 & 94.5 & 33.6\% \\
\midrule
\textbf{Overall} & \textbf{4.25\%} & \textbf{1.97\%} & \textbf{690,500} & \textbf{155.0} & \textbf{38.5\%} \\
\bottomrule
\end{tabular}
\vspace{-.01in}
\end{table*}
Versioned provenance is bimodal. Of the 1.97\% of events with a checkable version, 1.68\% are upstream-commit references concentrated in fork ecosystems (falling to 0.98\% once Linux/Chromium/Android projects are excluded), while only 0.28\% are independent version pins. Deliberate prose attribution (a developer stating in the commit message that a file was copied from a named source) appears in just 0.03\% of events, the floor of intentional documentation.

\textbf{Why low visibility implies high impact.} Visibility splits into two channels. Recorded provenance is \emph{inspectability}, not a freshness guarantee: documented copies are only marginally fresher than the rest (Cliff's $\delta=-0.03$, a negligible effect); its value is that a future maintainer can reconstruct origin when the gap must be acted on. Process visibility, by contrast, is strongly associated with maintenance state: PR-linked copies have median snapshot lag 2.0 vs.\ 175.4 days ($\delta=-0.41$), yet only 4.8\% of events are PR-linked.

The uninspectable gap footprint is large: 24.9\% of copy events (172,038) combine three or more gap signals with no recorded maintenance memory (no PR link, no documented provenance, no lifecycle text), and their median snapshot lag exceeds 1,400 days.

\vspace{-.05in}
\subsection{RQ3b: Maintenance}
\label{sec:rq3_implications}
\vspace{-.02in}
Within the same 690,500-event commit sample, copied sources have a conservative median snapshot lag of 155 days (95\% CI: [153.7, 156.7]), and 38.5\% (95\% CI: [38.4\%, 38.6\%]) of events copy source material whose freshest sampled blob is older than one year. The oldest sampled blob gives a harsher view: median age span 790.7 days, with 62.0\% including material over one year old. Snapshot lag measures freshness; age span measures audit burden. These are different gaps: a recent sync can still import a large historical batch requiring extensive review.

Lag is sharply form-specific (Table~\ref{tab:rq3a_rq3b}): the deliberate, own-history forms are freshest: project migration/refactoring (3 days) and upstream sync/forking (7 days) track active development, while course materials (269 days), scaffolding (229 days), and UI assets (218 days) carry the oldest snapshots. Dependency vendoring (107 days) is the analytically telling case: despite upstream bypass being its dominant rationale (so recent upstream material was copied), 31.4\% of vendored copies already exceed one year, because the vendor path freezes the snapshot until a maintainer manually updates.

The same PR-link effect seen in RQ3a holds for freshness: PR-linked copy events have median snapshot lag 2.0 days versus 175.4 days for direct commits, and median age span 99.8 versus 859.0 days, yet only 4.8\% of events are PR-linked.

Because snapshot lag is minted on a blob's exact byte content, any intermediate modification (re-vendoring, reformatting) resets it, so it is a conservative lower bound on the copied code's true age.

\textbf{Package-repository reach.} After de-noise filtering (Section~\ref{sec:rq3_method}), 2,556 ordinary unmanaged copies target 1,610 Repology-mapped package repositories. These copies are fresher than the de-noised non-package comparison set (median snapshot lag 85.7 vs.\ 167.0 days) and more often PR-linked (10.4\% vs.\ 4.3\% in that subset), yet 34.4\% already exceed one year of lag and few record a checkable upstream version; the version and origin are invisible to the package tooling that distributes them.

Maintenance also stratifies by RQ2 rationale boundary: Lineage copies (CF) have deep age span (610 days) but short snapshot lag (45 days) and low PR rates (7.5\%); Dependency copies (UB, OV) are freshest (5-day lag) and most reviewed (38.1\% PR-linked). Maintenance debt tracks rationale, not just form.

\vspace{-.05in}
\begin{summary-rq}
\textbf{RQ3a--b summary.} Copied sources rarely record a recoverable origin (4.3\% documented; 2.0\% with a checkable version), are rarely reviewed (4.8\% PR-linked),
and are stale (median 155 days; 38.5\% $>$1yr). PR-linked copies are far fresher (2 vs.\ 175 days), and the widest-gap forms (ad-hoc logic, data, course materials) are the least documented.
\end{summary-rq}
\vspace{-.05in}
\subsection{RQ3c: Security}
\label{sec:rq3c}
\vspace{-.02in}
For RQ3c we switch from the commit sample to the VFC-seeded dataset (Section~\ref{sec:rq3_method}): the CVE pipeline identifies 17,314 copy commits that copied blobs later associated with CVE-fixing commits. The security risk is highly concentrated by copy form: 15,163 of 17,314 CVE-risk copy commits (87.58\%) are dependency-vendoring cases, followed by upstream sync/forking (1,751; 10.11\%); all other forms together are under 2.5\%. Table~\ref{tab:cve_risk_surfaces} reports the full breakdown with repair observability bounds.

\begin{table*}[htbp]
\centering
\caption{Full-WoC CVE-risk copy commits by RQ1 copy form.
}
\label{tab:cve_risk_surfaces}
\vspace{-.1in}
\footnotesize
\begin{tabular}{lrrrrrrr c}
\toprule
\textbf{RQ1 copy form} & \textbf{Copies} & \textbf{Lower Fixed} & \textbf{Upper Fixed} & \textbf{Med. CVSS} & \textbf{Lower\%} & \textbf{Upper\%} & \textbf{Med. repair lag} & \textbf{Repair lag dist.} \\
\midrule
Dependency Vendoring & 15,163 & 6,699 & 12,635 & 7.5 & 44.18 & 83.33 & 677 days & \raisebox{-3pt}{\includegraphics[height=11pt]{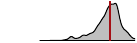}} \\
Upstream Sync \& Forking & 1,751 & 1,239 & 1,514 & 7.5 & 70.76 & 86.46 & 508 days & \raisebox{-3pt}{\includegraphics[height=11pt]{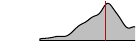}} \\
Other forms & 400 & 251 & 338 & 6.8 & 62.75 & 84.50 & 267 days & \raisebox{-3pt}{\includegraphics[height=11pt]{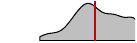}} \\
\midrule
\textbf{Overall} & \textbf{17,314} & \textbf{8,189} & \textbf{14,487} & \textbf{7.5} & \textbf{47.30} & \textbf{83.67} & \textbf{640 days} & \raisebox{-3pt}{\includegraphics[height=11pt]{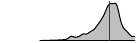}} \\
\bottomrule
\end{tabular}
\end{table*}
Among the 13,418 CVE-risk commits with CVSS metadata, 80.2\% score $\geq$7.0 (high) and 37.2\% score $\geq$9.0 (critical). Repair, when observable at all, is slow: across the 13,585 commits where both an upstream fix and a downstream edit-away are observable (Section~\ref{sec:rq3_method}), the median repair delay is 640~days (mean 769), with a long tail beyond three years (\emph{Repair dist.} column, Table~\ref{tab:cve_risk_surfaces}): copied vulnerabilities are not only severe but persistent.

Dependency-vendoring CVE copies have a much lower \emph{confirmed} repair rate than upstream-sync copies (44.18\% vs.\ 70.76\% lower bound), consistent with the maintenance gap: vendored copies lack the metadata that would surface an update need. Their permissive upper bounds are closer (83.33\% vs.\ 86.46\%): most vendored vulnerable files are eventually edited, but far fewer adopt the exact upstream fix. 

\textbf{Package-repository reach.} The security risk extends into packaged software: 65 project-CVE candidate pairs have Repology-mapped destinations, with exact fixed-blob adoption in 0 of 65 and only seven showing temporally plausible upper-bound replacements. The propagation path is visibly non-mechanical: a fix travels only when a maintainer notices, as the GDAL case (Section~\ref{sec:discussion}) illustrates.

\subsection{RQ3d: Compliance}
\label{sec:rq3d}

On the pruned project-pair license dataset (Section~\ref{sec:rq3_method}), the license pipeline identifies 41,777 pre-validation license-risk candidates across 3,687 source/destination project pairs and 97 incompatible license-pair combinations. The risk profile is the complement of the CVE profile: the source-code--family forms dominate (66.25\% of candidates), led by upstream sync/forking (51.49\%) with project migration (11.30\%) and ad-hoc logic (3.40\%), while dependency vendoring accounts for only 22.17\%. Like other source-code copies, these candidates seldom record a recoverable origin in commit text (Section~\ref{sec:rq3_attention}), so the license context typically crosses repository boundaries with little trace of where the file came from.

The automated matrix is deliberately broad, so we dual-audited the 326 high-star candidate rows ($\geq$1,000 destination stars; Section~\ref{sec:rq3_method}). A first human pass identified 60 likely violations (18.4\%; 95\% CI: [14.1\%, 22.7\%]); a second, independent pass then re-judged each candidate at the \emph{file level}, comparing the historically copied blob's own license (read from WoC) against the destination repository license, and treating documentation, manifests, and configuration as mere aggregation. This pass confirms 39 of the 326 candidates (12.0\%; 95\% CI: [8.6\%, 15.6\%])
as genuine incompatibilities, with binary inter-rater agreement of 93.6\% ($\kappa=0.752$); all 21 disagreements are one-directional, the file-level pass only downgrading first-pass positives. The large candidate-to-confirmed gap is the construct behaving as designed: repository-level licenses over-approximate file-level compatibility, and the audit corrects this by reading the copied file itself. Principal false-positive categories are the same third-party dependency in both projects (98), reversed direction (43), same author/owner (35), generated code (18), and dual licensing (16); confirmed cases span permissive-to-restrictive, restrictive-to-permissive, and restrictive-to-restrictive imports.

\textbf{Package-repository reach.} Of the 39 confirmed cases, 16 land in package-distributed destinations spanning 11 distinct projects; two of those destinations are themselves broadly packaged across $\geq$10 ecosystems: CodeLite (18) and Jellyfin (11), with WildFly (7) and SumatraPDF (6) next, and the Discourse/Mousetrap case (Section~\ref{sec:discussion}) is among this set. Together with the maintenance and security reach above, copy-transferred visibility gaps reach distributed software where no manifest records them.

\begin{summary-rq}
\textbf{RQ3c--d summary.} CVE risk concentrates in vendoring (88\%); across all CVE-risk commits, 80.2\% score CVSS $\geq$7.0 and exact-fix adoption is 47\%--84\%. License risk concentrates in source-code reuse (66\%; 39 confirmed cases, $\kappa=0.752$). Both reach packaged repositories where they are invisible to manifest-based tooling.
\end{summary-rq}

\vspace{-.08in}
\section{Discussion}
\label{sec:discussion}
\vspace{-.08in}

The results above are not simply a clone-management story. File-level copying changes the maintenance relationship around a file: who must track upstream changes, whether the source/version is recoverable, whether vulnerability fixes propagate, and whether legal context survives the move. Three audited cases make this concrete: \path{discourse} (2 ecosystems) vendors an Apache-2.0 file inside a GPLv2 repository with no manifest recording its origin; \path{codelite}~\cite{repo_codelite} (18 ecosystems) imports GPL-3.0 source into its GPL-2.0 codebase, a one-way copyleft incompatibility with no manifest recording it; and \path{gdal} (43 ecosystems) resynced its LibTIFF copy 3~days after the upstream CVE fix yet took 61~days to ship it to 20,768 dependent repositories (Section~\ref{chap:intro}). In every case, tooling operating on declared metadata alone sees nothing.

The pattern is not exceptional: 24.9\% of all copy events (172,038) carry three or more gap signals with no recorded maintenance memory. The mechanism is the same in every case: a copied file becomes the destination project's local responsibility without the version field, update channel, or manifest entry that would otherwise signal when that responsibility requires action. A copy is strictly less visible than a pinned package dependency (itself a documented maintenance burden~\cite{DBLP:journals/ese/KulaGOII18,DBLP:journals/cacm/Cox19}): there is no manifest entry to reveal that the snapshot has drifted, and no automated updater can propose a bump~\cite{DBLP:conf/icsr/ZeroualiCMRG18,DBLP:journals/smr/ZeroualiMGDCR19,chinthanet2021lags,DBLP:conf/icse/LiuCF00022}.

The central empirical finding is that copy form and rationale are associated but not interchangeable. Copy form is the observable triage signal (determining which visibility dimensions are \emph{affected}), while rationale explains which dimension \emph{dominates}; the two stages are complementary, not contradictory. Because rationale labels are recoverable for only 3,912 of 690,500 events, RQ3 uses form as the operational stratification variable; the form--rationale associations established in RQ2 then explain why each form concentrates a particular gap rather than distributing risk uniformly. The form-aware governance action each finding implies is stated as a bold \textbf{Design} requirement below. Maintenance (RQ3b) is cross-cutting rather than form-specific: a stale vendored copy is a security risk, a stale source copy is a license risk, and a stale fork copy is a traceability gap; each governance implication below presupposes the visibility and freshness conditions established in RQ3a--b.

A single ``no code clones'' rule is miscalibrated: it over-alerts on intentional fork synchronization and under-specifies the dependency and license gaps that matter. The form-aware responses follow.
\textbf{Copies are uninspectable and stale.} Only 4.3\% of copies document an origin and 2.0\% record a checkable version. \textbf{Design:} a pre-commit or PR hook that records the source URL, upstream commit SHA, and local-patch diff at copy time is the single most general intervention; PR-linked copies are already far fresher (median 2 vs.\ 175 days).

\textbf{Vendored copies are opaque dependency pins.} 88\% of CVE-risk copy commits reside in the dependency-vendoring form but carry no package manifest entry. 
\textbf{Design:} a blob-level CVE scanner that joins in-tree file SHAs against known vulnerable-blob tables surfaces the implicit security dependencies that manifest-based SCA misses.

\textbf{Source-code copies open a compliance gap.} Source-code reuse (66\% of license-risk candidates, chiefly fork/upstream sync) crosses license families without the attribution that a file header would preserve. 
\textbf{Design:} at pull-request time, a source-to-destination license-pair lookup flags cross-boundary imports and prompts the contributor for a license header, source URL, and compatibility justification before merge.

\textbf{Fork-maintenance and hardware copies require upstream traceability, not prohibition.} Hardware/driver copies (78\% patch porting among rationale-bearing commits) and cross-project patch porting are deliberate and necessary; the useful signal is upstream commit identity, not the copy itself. \textbf{Design:} commit-message linting that requires \texttt{Fixes:}, \texttt{Cherry-picked-from:}, or \texttt{Upstream-commit:} trailers, as already enforced in Linux kernel workflows, captures the provenance that our data shows is present in 46\% of CF-rationale commits but nearly absent in vendoring and source-code copies.

These tiers specify the requirements for an \emph{implicit dependency manager}: \emph{detect} copied blobs, \emph{classify} form, \emph{recover} origin and rationale, \emph{close} the dominant visibility gap with a form-specific policy, and \emph{emit} alerts. Every component except \emph{close} and \emph{emit} maps to an analysis method validated in this paper (Section~\ref{sec:data_construction}, 89.5\%; RQ1 and RQ2 labelers); the close policies and emit thresholds are the design-and-validation work a future tool must undertake. Such a tool would complement manifest-based SCA by operating on file content rather than declared metadata.

\section{Threats to Validity}
\label{sec:threats}
\vspace{-.08in}
\textbf{Construct validity.} All copy-form labels, both the RQ1--RQ2 association (Figure~\ref{fig:rq1_rq2_heatmap}) and the full-WoC RQ3c/d stratification (the 88\% CVE-in-vendoring and 66\% license-in-source concentrations), come from one \texttt{gemini-3-flash} labeler applying the 13-axial codebook to the same inputs (the copied file, its content, and the commit message), validated at $\kappa=0.845$ (human--human) and $\kappa=0.799$ (LLM--human) on 257 dual-annotated commits. RQ2 labels intent only where commit text gives sufficient evidence, so the 3,912 labeled commits are an intent-rich subset, not a prevalence sample (in-taxonomy agreement $\kappa=0.765$, raw 87.9\%, $n=231$); the main residual uncertainty is the in-scope/out-of-scope \textsc{EXCL} gate ($\kappa=0.650$ overall, $0.477$ binary), so we read rationale labels as aggregate evidence over in-scope codes, not as case-level or \textsc{EXCL}-prevalence claims. Our regexes capture \emph{recorded} provenance, not whether an origin exists.

\textbf{Internal validity.} Source-origin attribution can err when timestamps are skewed or histories mirrored; snapshot lag uses sampled blob-created times and is a conservative approximation of upstream state at copy time. The CVE repair bounds are intentionally wide (the lower bound misses manual patches; the upper bound can credit unrelated edits), and the license matrix is a repository-level candidate generator, so true compatibility depends on file-level licensing, linking context, and dual licensing, none of which it captures.

\textbf{External validity.} All analyses depend on WoC histories and mined copy relations, so missing repositories, private mirrors, and unindexed upstreams limit coverage. The OSV/CVE-linked ground-truth upstreams overrepresent security-sensitive, well-indexed projects, and the CVE analysis omits vulnerabilities without identifiable code-fixing commits (advisory-only or binary-only patches). The RQ3d compliance analysis is restricted to source/destination pairs both carrying $\geq$10 stars (dataset constraint~\cite{dabic2021sampling}), so the 39 confirmed violations are a lower bound biased toward prominent projects. Findings characterize open-source file-level copying observable in WoC and may not transfer to closed-source or non-Git ecosystems.

\section{Conclusion}
\label{sec:conclusion}
\vspace{-.10in}

File-level copying is a pervasive, ungoverned mode of reuse: across 690,500 copy events it spans 13 axial forms, from UI assets and scaffolding to vendored libraries and device drivers, each with a distinct visibility-gap profile. Observable form and developer rationale are associated but not interchangeable (hardware copies are 78\% fork-maintenance; vendored copies 70\% upstream-bypass), so form provides a practical triage signal even when commit-level rationale is unavailable at scale. The visibility loss is form-specific and reaches packaged software: CVE risk concentrates in dependency vendoring (88\%, 80\% high-severity) while license risk concentrates in source-code reuse (66\%), and both escape manifest-based tooling. These concentrations define the requirements for an implicit dependency manager: blob-level CVE scanning for vendored copies, license-pair checks for source-code reuse, and upstream-trailer linting for fork-maintenance copies.

\vspace{-0.18in}
\bibliographystyle{IEEEtran}
\bibliography{main}
\balance
\end{document}